\documentclass[pdflatex,sn-mathphys-num]{sn-jnl}

\usepackage{graphicx}%
\usepackage{multirow}%
\usepackage{amsmath,amssymb,amsfonts}%
\usepackage{amsthm}%
\usepackage{mathrsfs}%
\usepackage[title]{appendix}%
\usepackage{xcolor}%
\usepackage{textcomp}%
\usepackage{manyfoot}%
\usepackage{booktabs}%
\usepackage{algorithm}%
\usepackage{algorithmicx}%
\usepackage{algpseudocode}%
\usepackage{listings}%

\theoremstyle{thmstyleone}%

\theoremstyle{thmstyletwo}%

\theoremstyle{thmstylethree}%

\begin{document}

\title[Article Title]{Adaptive Shot Allocation for Recursive QAOA via Reinforcement Learning}


\author*[1,2]{\fnm{Euimin} \sur{Lee}}\email{euimin97@yonsei.ac.kr}

\author*[1,2]{\fnm{Shiho} \sur{Kim}}\email{shiho@yonsei.ac.kr}

\affil[1]{\orgdiv{School of Integrated Technology}, \orgname{Yonsei University}, \orgaddress{\postcode{03722}, \state{Seoul}, \country{Korea}}}

\affil[2]{\orgdiv{BK21 Graduate Program in Intelligent Semiconductor Technology}, \orgaddress{\orgname{Yonsei University}, \postcode{21983}, \state{Incheon}, \country{Korea}}}

\abstract{Recursive QAOA (RQAOA) solves combinatorial optimization problems by using shallow quantum circuits to estimate pairwise correlations and recursively eliminate variables until a classical solver can handle the residual instance. Each elimination step requires measurement shots, and the total shot cost grows with the number of recursive stages. On near-term quantum devices, increasing shot counts can translate directly into greater exposure to hardware-level noise sources such as readout errors and decoherence, making shot-efficient execution not merely a cost-reduction measure but a factor with direct implications for solution reliability. While shot reduction has been studied broadly across NISQ algorithms, step-wise measurement control inside the recursive loop of RQAOA has received little attention.
We formulate this step-wise allocation as a sequential decision problem and propose two complementary strategies for depth-1 RQAOA on weighted Max-Cut instances. A hand-crafted heuristic assigns shots based on local indicators of step difficulty, and a tabular Double Q-learning agent learns a residual policy that adjusts this baseline under a Lagrangian-constrained objective. Both methods are evaluated under a fixed-cap fairness protocol that equalizes the per-step budget across all strategies, and the elimination rule itself is kept unchanged so that the contribution of adaptive measurement control can be isolated.
On a diverse set of weighted graph instances spanning a range of sizes and structures, the heuristic reduces total shots by approximately 23\% relative to uniform allocation, and the RL policy achieves a 36\% reduction with a lower effective shots per success ratio than both baselines. The improvement persists on problem sizes not seen during training, suggesting that reinforcement learning can serve as an effective tool for discovering efficient, instance-adaptive measurement strategies in recursive quantum optimization.}


\maketitle

\section{Introduction}\label{sec1}
 
Quantum computing has opened new possibilities for optimization\cite{moll2018,cerezo2021}, simulation\cite{lee2025qwmc,bauer2020}, and learning\cite{biamonte2017} by exploiting quantum-mechanical features such as superposition, entanglement, and interference\cite{nielsen2010}. In the noisy intermediate-scale quantum (NISQ) regime\cite{preskill2018}, hybrid quantum--classical workflows---most prominently variational quantum algorithms\cite{peruzzo2014,mcclean2016}---have become a central design paradigm because they combine shallow circuits with classical outer-loop control rather than relying on full fault tolerance\cite{kandala2017,bharti2022,tilly2022}.
 
Among these methods, the Quantum Approximate Optimization Algorithm (QAOA) has attracted sustained attention as a variational approach tailored to combinatorial optimization\cite{farhi2014,hadfield2019,wang2018,zhou2020,crooks2018,willsch2020}. At the same time, its practical performance depends on circuit depth, parameter optimization, and the statistical quality of finite-shot expectation estimates. A substantial theoretical literature has also shown that low-depth QAOA can face locality \cite{farhi_typical2020,farhi_worst2020,marwaha2021,hastings2019} and trainability-induced limitations\cite{bravyi2020symmetry,mcclean2018,cerezo_barren2021,bittel2021} on important instance families. This motivates hybrid variants that retain shallow circuits while changing the algorithmic structure around them.
 
Recursive QAOA (RQAOA)\cite{bravyi2022graphcolor} is one such variant. Rather than attempting to solve the full optimization problem in one variational pass, RQAOA repeatedly estimates pairwise correlations from a shallow QAOA state, fixes a high-confidence relation between variables, contracts the instance, and then repeats the procedure on the reduced problem\cite{bae2023rqaoa,patel2024rlrqaoa,finzgar2024qiro,brady2024iterative}. This recursive structure can partially bypass some locality limitations of low-depth QAOA, but it introduces a different resource bottleneck: each elimination step requires measurement shots, and the cumulative cost can become substantial.
 
Because each elimination step requires its own set of circuit executions, the total shot cost compounds across the recursive chain. On near-term quantum devices, this cumulative measurement burden increases exposure to hardware-level noise sources such as readout errors and decoherence\cite{temme2017,endo2021}, making shot-efficient execution not merely a cost-reduction measure but a factor with direct implications for solution reliability. A broad literature has developed around shot-frugal variational optimization and adaptive gradient estimation\cite{kubler2020,arrasmith2020,gu2021,ito2023,sweke2020,stokes2020,spall1998}, operator grouping and measurement reduction\cite{zhao2021,huggins2021,gokhale2020,izmaylov2019,verteletskyi2020,crawford2021}, and AI driven shot reduction strategies\cite{liang2024}. However, most of this work addresses parameter-update loops in VQE or gradient-based settings. In RQAOA, measurements support discrete elimination decisions, and each decision changes the reduced instance that determines the next state. Recursion-level shot control is therefore a distinct sequential decision problem.

\begin{figure*}[t]
\centering
\includegraphics[width=\textwidth]{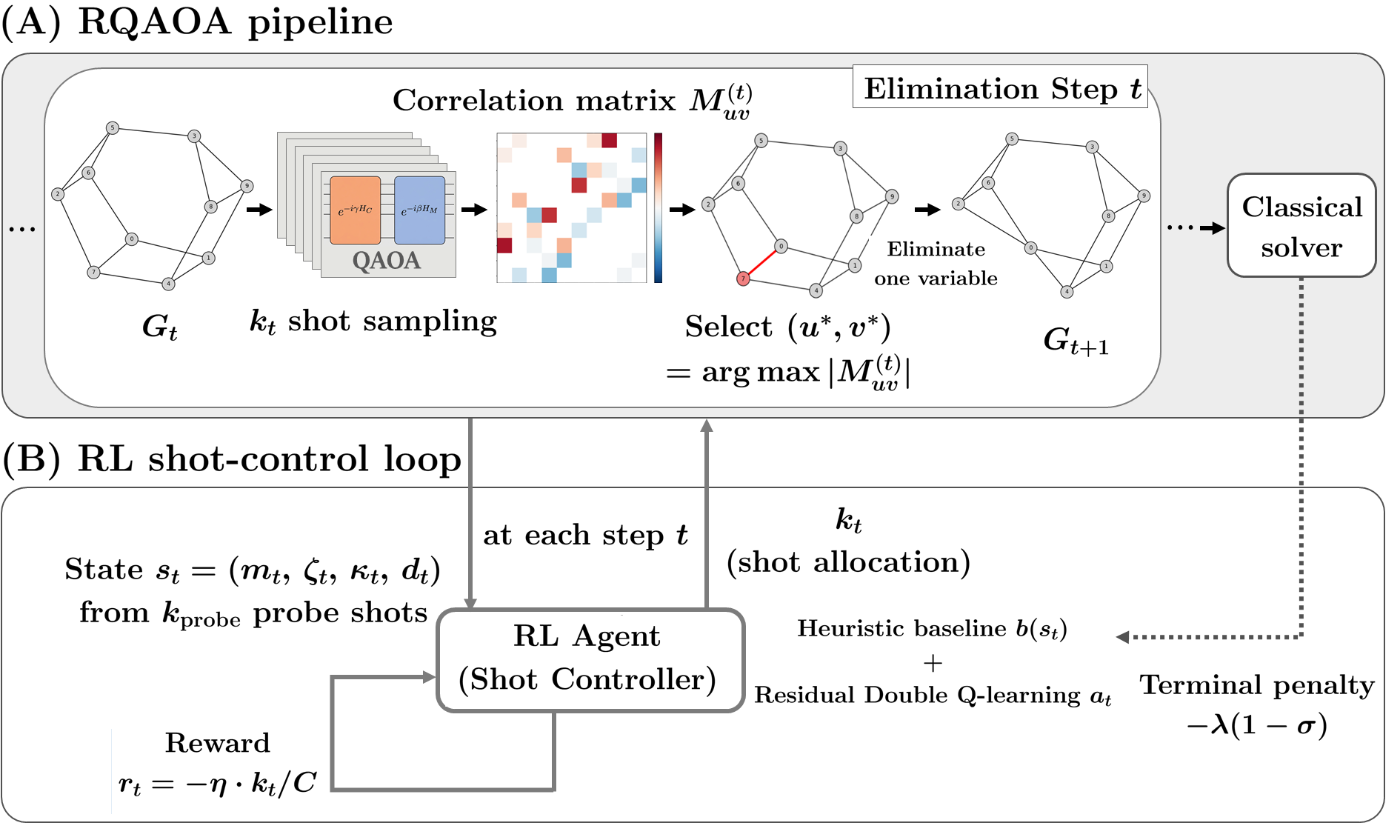}
\caption{Overview of the proposed shot-control framework for depth-1 RQAOA.
\textbf{(A)}~At each elimination step~$t$, a depth-1 QAOA circuit is executed on the current reduced graph~$G_t$ using $k_t$~measurement shots to estimate the pairwise correlation matrix~$M_{uv}^{(t)}$. The edge~$(u^{*}, v^{*})$ with the largest absolute correlation is selected, and the corresponding variable is eliminated to produce the next reduced graph~$G_{t+1}$. This procedure repeats until the problem size falls below a classical-solver threshold~$n_c$, at which point the residual instance is solved exactly.
\textbf{(B)}~The RL shot-control loop operates alongside the RQAOA pipeline. Before each elimination step, a small probe measurement of $k_{\mathrm{probe}}$~shots extracts a four-dimensional state descriptor $s_t = (m_t,\, \zeta_t,\, \kappa_t,\, d_t)$ that encodes the number of remaining variables, the z-gap between the two leading correlations, the conflict ratio among top candidate edges, and the graph distance between them. A tabular Double Q-learning agent takes this state as input and outputs a residual correction~$a_t$ to a hand-crafted heuristic baseline~$b(s_t)$, which together determine the shot allocation~$k_t$ for that step. The agent receives a per-step cost $r_t = -\eta \cdot k_t / C$ proportional to the fraction of the shot cap used, and a terminal penalty $-\lambda(1 - \sigma)$ that enforces a success-rate constraint via an adaptive Lagrangian multiplier.}
\label{fig:overview}
\end{figure*}
 
The key observation motivating this work is that not all elimination steps are equally demanding. When the leading pairwise correlation is well separated from the next candidate, even a modest number of shots suffices to identify the correct edge to contract. Conversely, when two or more correlations are nearly tied, substantially more shots may be needed to resolve the ambiguity. This variation in per-step difficulty suggests that a uniform shot budget across all steps is wasteful and that an allocation strategy responsive to the local correlation structure can reduce the total measurement cost without degrading solution quality.
 
We address this problem by proposing two allocation strategies. A state-aware Heuristic assigns shots based on local indicators of step difficulty, providing an interpretable rule that already achieves meaningful savings on its own. Separately, a tabular Double Q-learning agent learns a residual policy anchored to the Heuristic output, adjusting it where the fixed thresholds are too conservative or too aggressive. Both strategies keep the elimination rule fixed and control only the measurement budget, unlike methods that learn the elimination rule itself\cite{patel2024rlrqaoa} or the recursive structure\cite{finzgar2024qiro}. This separation isolates the contribution of adaptive shot allocation and allows the learned policy to be reused across instances without retraining the solver logic, building on the broader body of RL-based quantum optimization\cite{wauters2020,yao2020,khairy2020} and circuit-design work\cite{ostaszewski2021,skolik2023,altmann2024}.
 
The contributions of this work are fourfold. First, we formulate step-wise shot allocation in depth-1 RQAOA as a measurement-control problem over the recursive chain. Second, we propose a state-aware Heuristic that provides an interpretable standalone allocation rule, together with a tabular residual Double Q-learning controller that refines its output. Third, on held-out test instances, the Heuristic reduces shot usage by approximately 23\% relative to Uniform allocation, and the RL policy achieves about 36\% reduction with a lower effective shots per success than both baselines. Fourth, the learned policy captures transferable step-level corrections that generalize to unseen problem sizes within the hard $d$-regular weighted Max-Cut family, indicating that the state representation encodes local step difficulty rather than a narrow training-size rule. 
 
The remainder of this paper is organized as follows. Section~\ref{sec:background} reviews the problem setting and related prior work. Section~\ref{sec:method} presents the proposed shot-control framework. Section~\ref{sec:experiments} describes the experimental setup. Section~\ref{sec:results} reports the evaluation results, and Section~\ref{sec:discussion} discusses interpretation and limitations. Finally, Section~\ref{sec:conclusion} concludes.

\section{Background}
\label{sec:background}

\subsection{Weighted Max-Cut and Ising formulation}
\label{subsec:problem_setting}
 
We consider the weighted Max-Cut problem on a graph $G=(V,E,w)$. Let each vertex $u \in V$ be associated with a binary spin variable $z_u \in \{-1,+1\}$. Up to an additive constant, the weighted Max-Cut objective can be written in Ising form as
\begin{equation}
H_C = \sum_{(u,v)\in E} J_{uv} Z_u Z_v,
\label{eq:cost_hamiltonian}
\end{equation}
where $Z_u$ denotes the Pauli-$Z$ operator associated with vertex $u$, and $J_{uv}$ is the coefficient induced by the edge weight\cite{lucas2014,goemans1995}. Max-Cut is NP-hard in general, and the best known polynomial-time approximation ratio is approximately $0.878$\cite{goemans1995}. In the weighted setting, the heterogeneity of edge coefficients $J_{uv}$ amplifies the sensitivity of each elimination decision in recursive solvers, since a single misidentified correlation can propagate through edges of very different magnitudes. In this work, we focus on the weighted Ising setting without external field terms; the goal is to identify low-energy configurations of $H_C$, which correspond to high-quality cuts of the graph.
 
\subsection{Depth-1 QAOA and recursive reduction in RQAOA}
\label{subsec:qaoa_rqaoa}
 
QAOA prepares a variational quantum state by alternating applications of the cost Hamiltonian $H_C$ and a mixer Hamiltonian
\begin{equation}
H_M = \sum_{u\in V} X_u.
\label{eq:mixer_hamiltonian}
\end{equation}
A depth-$p$ QAOA state is
\begin{equation}
|\psi_p(\boldsymbol{\gamma}, \boldsymbol{\beta})\rangle
=
\prod_{k=1}^{p}
e^{-i\beta_k H_M}
e^{-i\gamma_k H_C}
|+\rangle^{\otimes n}.
\label{eq:qaoa_state}
\end{equation}
In the present work we focus on the $p=1$ setting. Depth-1 expectation values admit efficient classical simulation via closed-form expressions for the Ising cost function\cite{bravyi2022graphcolor,ozaeta2022}, enabling large-scale benchmarking that would be infeasible at higher depths, and this restriction provides a controlled testbed for isolating recursion-level measurement effects from confounding factors such as deeper-circuit trainability\cite{bravyi2022graphcolor,patel2024rlrqaoa}. Accordingly, we intentionally restrict attention here to the noiseless $p=1$ regime so that recursion-level shot allocation can be analyzed cleanly, without conflating it with hardware-noise effects or the additional performance variability introduced by optimizing deeper QAOA subroutines; these extensions are left to future work.
 
RQAOA uses this shallow QAOA state not as a one-shot solver, but as a subroutine inside a recursive elimination loop. Given the current reduced Hamiltonian $H_t$, we prepare $|\psi_1(\gamma_t,\beta_t)\rangle$ and estimate the pairwise correlations
\begin{equation}
M_{uv}^{(t)}
=
\langle \psi_1(\gamma_t,\beta_t) | Z_u Z_v | \psi_1(\gamma_t,\beta_t) \rangle .
\label{eq:pairwise_correlation}
\end{equation}
We then select
\begin{equation}
(u^\ast, v^\ast)
=
\arg\max_{(u,v)\in E_t}
\left| M_{uv}^{(t)} \right|,
\label{eq:max_correlation}
\end{equation}
and impose the constraint
\begin{equation}
Z_{u^\ast}
=
\mathrm{sign}\!\left(M_{u^\ast v^\ast}^{(t)}\right)\, Z_{v^\ast},
\label{eq:elimination_constraint}
\end{equation}
thereby eliminating one variable and producing a reduced Hamiltonian $H_{t+1}$. This procedure is repeated until the number of remaining variables falls below a predefined threshold $n_c$, after which the residual problem is solved classically\cite{bravyi2022graphcolor,bae2023rqaoa,patel2024rlrqaoa}.
 
\subsection{Related work}
\label{subsec:related_work}
 
Finite-shot estimation is a central practical bottleneck in hybrid quantum algorithms. Since expectation values must be inferred from repeated circuit executions, higher precision generally demands more shots and therefore greater execution time and hardware cost\cite{mcclean2016,cerezo2021,bharti2022}. A substantial literature has accordingly developed on shot reduction, adaptive shot use, operator grouping, and measurement optimization. Within the gradient-based optimizer family, adaptive shot heuristics such as iCANS, gCANS, and Rosalin adjust the per-iteration shot budget to reduce total measurement cost\cite{kubler2020,arrasmith2020,gu2021}; stochastic gradient and simultaneous-perturbation methods further reduce per-step overhead\cite{sweke2020,stokes2020,spall1998}; and latency-aware scheduling explicitly accounts for hardware turnaround time\cite{ito2023}. A parallel line of work addresses the measurement problem from the operator side through Pauli grouping\cite{zhao2021,huggins2021}, commuting-family partitioning\cite{gokhale2020,izmaylov2019}, and clique-cover or finite-sampling-aware strategies\cite{verteletskyi2020,crawford2021}. In the VQE domain, Liang \emph{et al.} use RL to dynamically determine the total shot count across optimization iterations, demonstrating transferability across molecular systems\cite{liang2024}. These efforts have been concentrated in VQE and gradient-based variational settings, and to our knowledge no prior work has addressed step-wise measurement-budget control within the recursive elimination chain of RQAOA.

Within the recursive quantum optimization literature, Bravyi \emph{et al.} 
formalize hybrid recursive reduction for graph coloring\cite{bravyi2022graphcolor}, 
and subsequent work has extended recursive quantum optimization to complete 
graphs\cite{bae2023rqaoa}, quantum-informed recursive 
optimization\cite{finzgar2024qiro}, and maximum independent 
set\cite{brady2024iterative}. Patel \emph{et al.} learn the elimination 
policy itself using REINFORCE\cite{patel2024rlrqaoa}. None of these works, 
however, study how the measurement budget should be allocated across 
elimination steps.

Reinforcement learning has proven effective for sequential resource-allocation 
problems in quantum computing more broadly, including QAOA parameter 
optimization\cite{wauters2020,yao2020}, ansatz and circuit 
design\cite{ostaszewski2021,kuo2021,fosel2021}, and robust control under 
noise\cite{fosel2018,niu2019,bukov2018,skolik2023}; recent surveys can be 
found in \cite{altmann2024,khairy2020}. Building on this line, we apply RL 
to the complementary problem of recursion-level shot allocation: the 
elimination rule is kept fixed, and the agent learns only how many shots to 
spend at each step, so that the effect of adaptive measurement control can be 
evaluated independently of changes to the solver itself.

\section{Shot Control Framework for RQAOA}
\label{sec:method}
 
Each contraction in RQAOA alters the reduced graph and reshuffles the ranking of candidate edges, so the correlation structure that determines step difficulty is different at every elimination step. A fixed allocation rule cannot account for this variation and will inevitably over-allocate shots on easy steps or under-allocate on hard ones. This motivates a state-dependent allocation strategy that reads the local difficulty of each step and sets the measurement budget accordingly.
 
We formulate step-wise shot allocation in depth-1 RQAOA as a finite-horizon Markov decision process (MDP) whose episode corresponds to a single RQAOA run. At each elimination step the agent observes the current state of the reduced instance, allocates a measurement budget, and receives a reward signal that penalizes shot usage subject to a success constraint. The controller follows a two-level design: a hand-crafted Heuristic baseline encodes simple domain rules about when a step is likely to be easy or hard, and a residual RL policy learns step-by-step corrections to that baseline. The core design choices are therefore a lightweight tabular state, a baseline-guided residual action space, and a Lagrangian-constrained training objective\cite{sutton2018,watkins1992,hasselt2010,altman1999,achiam2017,tessler2019}.
 
\subsection{MDP formulation}
\label{sec:mdp}
 
Let $n$ denote the initial number of variables and $n_c$ the classical-solver threshold. One RQAOA episode consists of $T = n - n_c$ elimination steps, indexed $t = 1, \ldots, T$. At step~$t$, the agent observes a state $s_t$ summarizing the reduced instance with $m_t$ active variables, selects a shot allocation $k_t \leq C$ (where $C$ is a per-instance shot cap), and observes a contraction outcome. The instance then transitions to $m_{t+1} = m_t - 1$. When $m_T = n_c$ the residual problem is solved classically and the episode terminates with a binary success indicator:
\begin{equation}
  \sigma = \mathbf{1}\!\left[\frac{E_{\mathrm{out}}}{E_{\mathrm{opt}}}
  \geq \rho^*\right],
  \label{eq:success}
\end{equation}
where $E_{\mathrm{opt}}$ is the brute-force optimum and $\rho^* = 0.99$.
 
\subsection{State representation}
\label{sec:state}
 
A lightweight four-dimensional state captures the difficulty of the current elimination step:
\begin{equation}
  s_t = \bigl(m_t,\; \zeta_t,\; \kappa_t,\; d_t\bigr),
  \label{eq:state}
\end{equation}
where $m_t \in [n_c{+}1,\, n]$ is the number of remaining variables, encoding position within the recursive chain. The \emph{z-gap}
\begin{equation}
  \zeta_t = \frac{|M_{(1)}^{(t)}|}{|M_{(2)}^{(t)}| + \varepsilon},
  \qquad \varepsilon = 10^{-12},
  \label{eq:zgap}
\end{equation}
is the ratio of the two largest absolute pairwise correlations; $\zeta_t \gg 1$ indicates that the leading edge is easily identifiable even with few shots, while $\zeta_t \approx 1$ signals a near-tie vulnerable to sampling noise. The \emph{conflict ratio} measures endpoint overlap among the top-$k$ candidate edges ($k=3$). Let $\mathcal{V}_k^{(t)}$ denote the set of unique endpoints of these $k$ edges. Then
\begin{equation}
  \kappa_t = 1 - \frac{|\mathcal{V}_k^{(t)}|}{2k}.
  \label{eq:conflict}
\end{equation}
High $\kappa_t$ means that eliminating one variable directly affects the next candidates, amplifying the downstream cost of a misidentified leading edge. Finally, the \emph{edge distance}
\begin{equation}
  d_t = \min_{u \in e_1,\, v \in e_2}
  \mathrm{dist}_G(u,v),
  \label{eq:dist}
\end{equation}
is the minimum graph-theoretic distance between the endpoints of the two most strongly correlated edges $e_1, e_2$; small $d_t$ indicates shared neighborhood structure.
 
State features are estimated from a probe measurement of $k_{\mathrm{probe}} \in [16,\,32]$ shots performed before the main allocation; all reported shot counts include these probe shots. Each feature is discretized into a small number of bins ($m_t$: $n{-}n_c$ levels; $\zeta_t$: 7 levels; $\kappa_t$: 5 levels; $d_t$: 5 levels), yielding a compact tabular state space.
 
\subsection{Action space and residual policy}
\label{sec:action}
 
Rather than selecting an absolute shot fraction from scratch, the controller begins from a hand-crafted Heuristic baseline and learns only residual corrections around it. This baseline is not merely an initialization device for RL; it is a standalone, state-aware allocation rule designed from local indicators of step difficulty. Six candidate fractions of the cap $C$ are available,
$\mathcal{F} = \{0.20,\, 0.35,\, 0.50,\, 0.65,\, 0.80,\, 1.00\}$,
indexed $0,\ldots,5$. The baseline index $b(s_t)$ assigns an initial allocation level based on the state features:
\begin{equation}
  b(s_t) = \begin{cases}
    0 & \zeta_t \geq 4.0,\; \kappa_t < 0.10,\; d_t \geq 3
      \quad (\text{use }20\%\text{ of }C), \\
    1 & \zeta_t \geq 2.0,\; \kappa_t < 0.20,\; d_t \geq 2
      \quad (\text{use }35\%), \\
    4 & \zeta_t < 0.9 \text{ and } (\kappa_t \geq 0.30 \text{ or } d_t \leq 1)
      \quad (\text{use }80\%), \\
    2 & \text{otherwise}
      \quad (\text{default, }50\%).
  \end{cases}
  \label{eq:baseline}
\end{equation}
 
We evaluate this hand-crafted rule throughout as the standalone Heuristic baseline. This makes it possible to separate two questions experimentally: how far an interpretable domain-informed policy can go on its own, and how much additional benefit is obtained when RL learns state-dependent corrections to that policy.
 
The agent's action $a_t \in \mathcal{A} = \{-3,-2,-1,0,{+}1,{+}2\}$ is a residual offset applied to this baseline:
\begin{equation}
  f_t = \mathrm{clip}\bigl(b(s_t) + a_t,\; 0,\; 5\bigr),
  \qquad
  k_t = \max\!\bigl(k_{\mathrm{probe}},\;
  \lfloor \mathcal{F}[f_t] \cdot C \rceil\bigr).
  \label{eq:action}
\end{equation}
 
This residual design concentrates exploration near an interpretable and already competitive prior rather than sweeping over the full action range. It also guarantees that the hand-crafted rule remains exactly available through $a_t = 0$. The role of RL is therefore not to replace a weak comparator, but to learn when the baseline is too conservative or too aggressive, especially in regimes where fixed thresholds do not transfer perfectly across problem sizes.
 
\subsection{Reward and Lagrangian-constrained objective}
\label{sec:reward}
 
Per-step rewards penalize shot expenditure relative to the cap:
\begin{equation}
  r_t = -\eta \cdot \frac{k_t}{C}, \qquad t = 1,\ldots,T,
  \label{eq:step_reward}
\end{equation}
with $\eta = 1.0$. At the terminal step, an additional failure penalty is incurred if the episode does not meet the success criterion:
\begin{equation}
  r_T \;\mathrel{+}=\; -\lambda \cdot (1 - \sigma),
  \label{eq:terminal_reward}
\end{equation}
where $\sigma$ is the success indicator defined in Eq.~\eqref{eq:success} and $\lambda \geq 0$ is a Lagrangian multiplier.
 
The multiplier enforces the constraint
$\mathbb{P}_\pi[\sigma = 1] \geq p^*$ with target $p^* = 0.95$. It is adapted across episodes via
\begin{equation}
  \lambda \leftarrow \mathrm{clip}\!\bigl(
    \lambda + \mu_\lambda \cdot (p^* - \hat{p}),\;
    0,\; \lambda_{\max}
  \bigr),
  \label{eq:lambda_update}
\end{equation}
where $\hat{p}$ is an exponential moving average of episode-success indicators with coefficient $\beta = 0.10$, and $\lambda_{\max} = 80$. During a warm-up phase of 100 episodes, $\lambda$ is held at its initial value $\lambda_0 = 2.0$ to allow the policy to discover successful trajectories before the constraint tightens.
 
The overall training objective is
\begin{equation}
  \min_\pi\;\mathbb{E}_\pi\!\left[\sum_{t=1}^{T} k_t\right]
  \quad \text{subject to} \quad
  \mathbb{P}_\pi[\sigma = 1] \geq p^*.
  \label{eq:objective}
\end{equation}
 
\subsection{Double Q-learning algorithm}
\label{sec:training}
 
We maintain two independent tabular Q-function estimates $Q_1, Q_2 :
\mathcal{S} \times \mathcal{A} \to \mathbb{R}$, trained via Double Q-learning\cite{hasselt2010} to mitigate maximization bias. At each update, one table is selected uniformly at random; $Q_1$ is updated using the greedy action from $Q_2$ as the bootstrap target and vice versa:
\begin{equation}
  Q_i(s,a) \leftarrow (1{-}\alpha)\,Q_i(s,a) + \alpha\bigl[
    r + \gamma\, Q_j\!\bigl(s',\,\arg\max_{a'} Q_i(s',a')\bigr)
  \bigr],
  \label{eq:doubleq}
\end{equation}
where $(i,j)$ is a random permutation of $(1,2)$.

\begin{algorithm}[t]
\caption{Lagrangian-constrained residual Double Q-learning for RQAOA shot allocation}
\label{alg:training}
\begin{algorithmic}[1]
\Require Instance $(\mathbf{J}, E_{\mathrm{opt}}, C, n, n_c)$; hyperparameters $\alpha, \gamma, \lambda_0, \lambda_{\max}, p^*$
\State Initialize $Q_1, Q_2 \leftarrow \mathbf{0}$,\;
       $\lambda \leftarrow \lambda_0$,\;
       $\hat{p} \leftarrow p^*$,\;
       $\epsilon \leftarrow 1.0$
\For{episode $e = 1,\ldots,1200$}
  \State $m \leftarrow n$;\; $\mathbf{J}' \leftarrow \mathbf{J}$
  \While{$m > n_c$}
    \State Probe with $k_{\mathrm{probe}}$ shots and extract $(\zeta,\,\kappa,\,d)$
    \State $s \leftarrow \mathrm{encode}(m,\,\zeta,\,\kappa,\,d)$;\;
           $b \leftarrow \mathrm{baseline}(\zeta,\,\kappa,\,d)$
    \State $a \leftarrow \epsilon$-greedy$(Q_1{+}Q_2,\, s)$;\;
           $f \leftarrow \mathrm{clip}(b{+}a,\,0,\,5)$
    \State $k \leftarrow \max\!\bigl(k_{\mathrm{probe}},\,
           \lfloor \mathcal{F}[f] \cdot C \rceil\bigr)$
    \State Execute RQAOA contraction with $k$ shots;\;
           $r \leftarrow -k/C$;\;
           $m \leftarrow m{-}1$
    \State Update $Q_1$ or $Q_2$ via Eq.~\eqref{eq:doubleq}
  \EndWhile
  \State Solve the residual problem classically and compute $\sigma$ via Eq.~\eqref{eq:success}
  \State $r_T \mathrel{+}= -\lambda\,(1{-}\sigma)$;\; update terminal Q-value
  \If{$e > 100$}
    \State $\hat{p} \leftarrow (1{-}\beta)\hat{p} + \beta\,\sigma$;\; update $\lambda$ via Eq.~\eqref{eq:lambda_update}
  \EndIf
  \State $\epsilon \leftarrow \max(0.02,\;\epsilon \times 0.995)$
\EndFor
\State \Return best-checkpoint policy
\end{algorithmic}
\end{algorithm}

An episode proceeds as follows. At each step~$t$: (i) a probe measurement of $k_{\mathrm{probe}}$ shots is performed to compute $\zeta_t$, $\kappa_t$, and $d_t$; (ii) the state $s_t$ is encoded and discretized; (iii) the $\epsilon$-greedy policy selects $a_t$ and yields $k_t$ total shots for the step; (iv) the contraction is executed; and (v) the transition $(s_t, a_t, r_t, s_{t+1})$ is used immediately to update $Q_1$ or $Q_2$.
 
Hyperparameters are fixed across all training runs: learning rate $\alpha = 0.15$, discount $\gamma = 0.97$, $\epsilon$-decay from $1.0$ to $0.02$ at rate $0.995$ per episode, and $1{,}200$ training episodes for each base-policy run. Each base policy is trained only on its designated training instance, and the learned policy is then evaluated, without further updating, on held-out and cross-size instances. Checkpoint selection retains the policy with the highest validation success rate, breaking ties by lower median total shots and then lower mean total shots.

\section{Experimental Setup}
\label{sec:experiments}\label{sec:experimental_setup}
 
All experiments are conducted on a noiseless simulator using Qiskit Aer (\texttt{AerSimulator} backend)\cite{aleksandrowicz2019}. Full implementation details, including angle optimization, probe-shot settings, and RL hyperparameters, are provided in Appendix~\ref{app:impl}.
 
\subsection{Training bases and evaluation pools}
\label{sec:instances}
 
We evaluate two base policies, each trained on one designated weighted Max-Cut instance: \textbf{Base~0} ($b_0$), an 8-regular graph on $n = 14$ nodes, and \textbf{Base~1} ($b_1$), a 17-regular graph on $n = 20$ nodes. AAll graphs in the benchmark are $d$-regular with Gaussian edge weights. The benchmark combines designated hard training bases, reweighted variants of the base topologies, additional same-size hard instances from the pool, same-size random regular graphs without the hard-instance filter, and cross-size instances drawn from the hard-instance pool. Here, ``hard'' is used operationally only for instances whose Uniform-allocation approximation ratio is at most 0.95 relative to the brute-force optimum. All experiments use the same classical threshold $n_c = 8$ and circuit depth $p = 1$.
 
Each base policy is evaluated on a dedicated set of same-size test instances (reweighted variants of the training graph, different hard instances from the pool, and randomly generated regular graphs) as well as a shared \textbf{Cross-size} set spanning unseen node counts $n \in \{15,\ldots,25\}\setminus\{14,20\}$. The complete benchmark contains \textbf{104 unique graph instances} and \textbf{164 policy--instance evaluation pairs}.
 
\subsection{Cap calibration and evaluation protocol}
\label{sec:cap}
 
For each instance, the per-step shot cap $C_{\mathrm{uniform}}$ is determined using the Uniform baseline only. A two-stage search identifies the smallest cap such that Uniform achieves success rate $\geq 0.95$ over $N_{\mathrm{cal}} = 60$ calibration trials. The search first scans a coarse power-of-two grid $\{64, 128, \ldots, 4096\}$ and then refines the result within the identified bracket. All three methods are subsequently evaluated under this same cap, so that any observed difference in shot usage or success rate reflects the allocation strategy rather than unequal per-step budgets.
 
Some evaluation pairs remain strongly budget-limited even at the maximum cap. We retain only pairs for which $\mathrm{SR}_{\mathrm{Uniform}} \geq 0.90$, yielding an \textbf{Operational subset} of 135 pairs. Excluding the Training and Validation categories used for model selection leaves the \textbf{held-out operational subset} of 130 pairs over 81 unique instances, which is the basis for all main quantitative claims. The remaining 29 pairs are not discarded but reported separately as a robustness view (Complete benchmark, 164 pairs).
 
\subsection{Baselines and metrics}
\label{sec:baselines}
 
We compare three allocation strategies, each evaluated over $N = 60$ independent trials per policy--instance pair. \textbf{Uniform} allocates exactly $C_{\mathrm{uniform}}$ shots at every elimination step. \textbf{Heuristic} applies the hand-crafted baseline $b(s_t)$ from Eq.~\eqref{eq:baseline} without RL adjustment ($a_t = 0$ throughout), serving both as a standalone policy and as the anchor for the residual RL controller. \textbf{RL} ($b_0$ or $b_1$) applies the learned Double Q-policy trained on the designated base instance.
 
For each pair, \emph{shot reduction} is defined as
\begin{equation}
  \text{Reduction} = 1 - \frac{\widetilde{S}_{\text{method}}}{\widetilde{S}_{\text{Uniform}}},
  \label{eq:reduction}
\end{equation}
where $\widetilde{S}$ denotes the median total shots over all trials. \emph{Success rate} (SR) is the fraction of trials satisfying $E_{\mathrm{out}} / E_{\mathrm{opt}} \geq 0.99$. To combine resource usage and reliability into a single efficiency proxy, we define the \emph{effective shots per success}:
\begin{equation}
  \mathrm{ESP} = \frac{\widetilde{S}_{\mathrm{succ}}}{\mathrm{SR}},
  \label{eq:esp}
\end{equation}
where $\widetilde{S}_{\mathrm{succ}}$ is the median total shots among successful trials. Lower $\mathrm{ESP}$ indicates better efficiency. For cross-method comparison we report the normalized ratio $\mathrm{ESP}_{\text{method}} / \mathrm{ESP}_{\text{Uniform}}$, for which values below 1.0 indicate improvement over Uniform. During training, the policy snapshot with the highest validation success rate is retained, breaking ties by lower median total shots.
 
\section{Results}
\label{sec:results}

All three methods are evaluated under the fixed-cap fairness protocol of Sec.~\ref{sec:cap}. The main quantitative claims below are based on the \textbf{held-out operational subset} (130 pairs, 81 unique instances) that excludes Training and Validation categories.

\subsection{Overall and regime-level performance}
\label{sec:results_aggregate}

Table~\ref{tab:aggregate} summarizes aggregate performance across the main held-out test set and robustness subsets.

\begin{table}[t]
\centering
\caption{Aggregate performance on the main held-out test set and robustness subsets. Reduction and ESP ratio are measured relative to Uniform.}
\label{tab:aggregate}
\small
\begin{tabular}{@{}lrrrrr@{}}
\toprule
Subset & $N$ & RL Red. & RL ESP & Heu.\ Red. & Heu.\ ESP \\
\midrule
Held-out operational pairs & 130 & 36.1\% & 0.676 & 23.0\% & 0.790 \\
\;\;Same-size held-out      &  32 & 34.8\% & 0.684 & 25.3\% & 0.766 \\
\;\;Cross-size pairs               & 98 & 36.6\% & 0.673 & 22.2\% & 0.798 \\
\midrule
Operational pairs (incl.\ Train/Val) & 135 & 36.1\% & 0.677 & 23.0\% & 0.790 \\
Complete benchmark           & 164 & 36.5\% & 0.674 & 22.3\% & 0.798 \\
\bottomrule
\end{tabular}
\end{table}

On the held-out operational subset, the RL policy achieves a mean pairwise shot reduction of \textbf{36.1\%} relative to Uniform, compared with \textbf{23.0\%} for the Heuristic. The corresponding ESP ratios are \textbf{0.676} (RL) and \textbf{0.790} (Heuristic), indicating that RL attains a substantially lower effective cost per success. Including Training/Validation instances or the full Complete benchmark changes these figures only negligibly, confirming that the qualitative ranking is not driven by the filtering rule.

Breaking the held-out subset into evaluation regimes, RL remains effective in both: same-size held-out (34.8\% reduction, 0.684 ESP ratio) and Cross-size (36.6\%, 0.673). The gain is at least as strong on unseen sizes. The Heuristic's strong same-size performance (25.3\% reduction) is itself informative: the hand-crafted rule captures substantial allocation structure without any learning. The RL--Heuristic gap varies with instance complexity, being modest at $n{=}14$ (only six elimination steps) and consistently larger at $n{=}20$ and beyond.

\subsection{Base policy comparison}
\label{sec:results_base}

The benchmark contains 51 \emph{common} instances from the Operational subset that are evaluated under both $b_0$ and $b_1$. These all belong to the \textbf{Cross-size} regime, so the comparison isolates transfer quality on unseen sizes.

\begin{table}[t]
\centering
\caption{Comparison of the two RL base policies on the 51 matched cross-size instances evaluated under both policies.}
\label{tab:base}
\small
\begin{tabular}{@{}lrrrr@{}}
\toprule
Policy & $N$ & Reduction & ESP ratio & Success rate \\
\midrule
$b_0$ (trained at $n=14$) & 51 & 34.7\% & 0.691 & 0.915 \\
$b_1$ (trained at $n=20$) & 51 & 38.4\% & 0.655 & 0.911 \\
\bottomrule
\end{tabular}
\end{table}

On this matched subset, $b_1$ achieves both higher reduction and lower ESP ratio than $b_0$, while the two policies have very similar success rates. One plausible explanation is that training on the larger $n=20$ base exposes the agent to a broader range of elimination-step difficulties.

\subsection{Generalization to unseen problem sizes}
\label{sec:results_cross}

The Cross-size regime provides the most demanding transfer test, spanning $n \in \{15,\ldots,25\}\setminus\{14,20\}$ with 98 policy--instance pairs. The mean RL shot reduction on these unseen-size instances is \textbf{36.6\%}, essentially matching the overall held-out average. Within the $d$-regular Gaussian-weighted benchmark family used in this study, this persistence suggests that the state representation captures local step difficulty in a transferable form rather than memorizing size-specific thresholds.

\begin{figure}[t]
  \centering
  \includegraphics[width=0.85\linewidth]{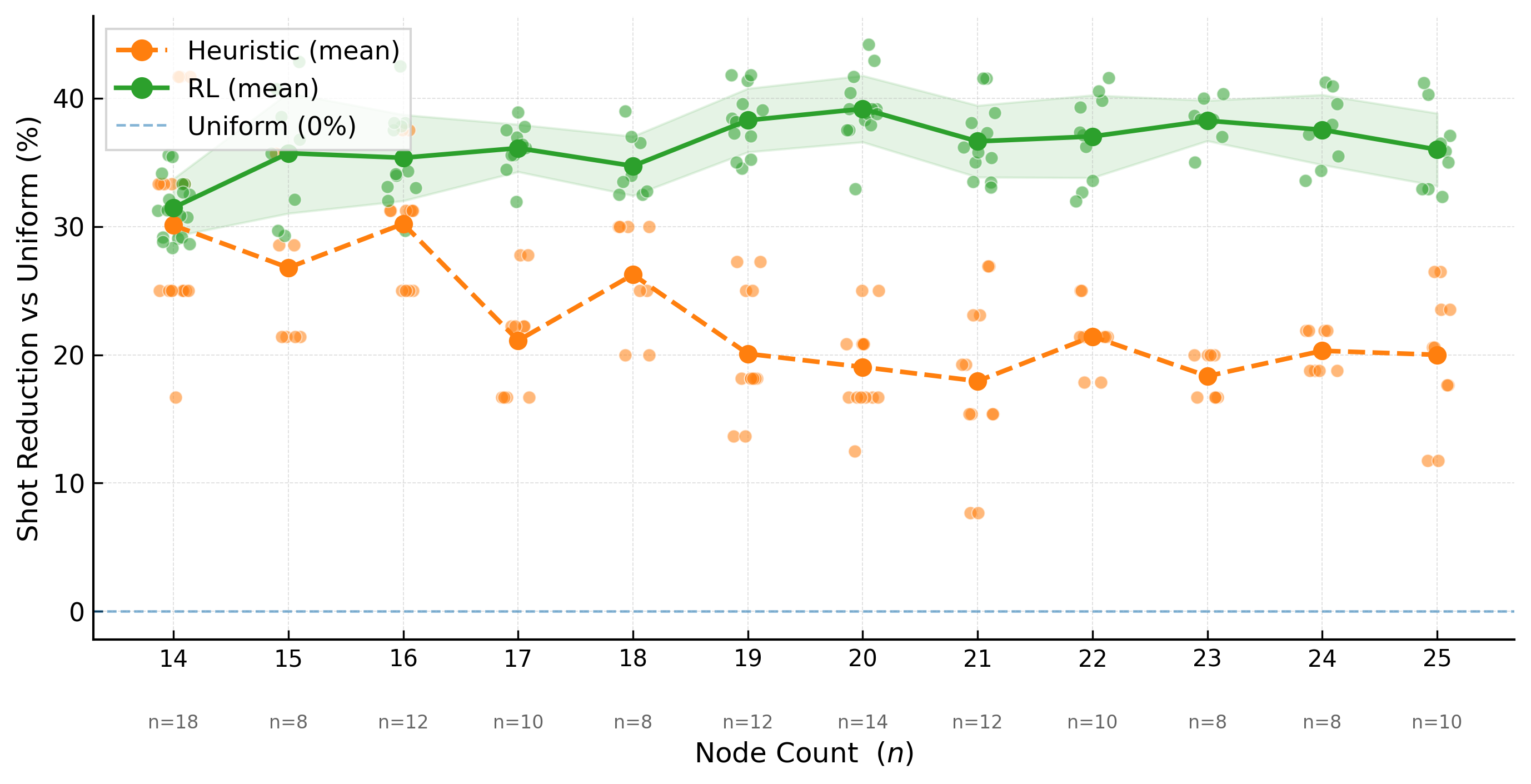}
  \caption{Shot reduction (\%) vs.\ problem size $n$ for the held-out operational subset. Each point represents one policy--instance pair. The per-size mean (connected line) shows that RL achieves stable ${\sim}36\%$ reduction across the full range, while the Heuristic settles around ${\sim}23\%$.}
  \label{fig:reduction_vs_n}
\end{figure}

Figure~\ref{fig:reduction_vs_n} shows the shot reduction as a function of problem size, confirming that the RL advantage is stable across the full $n$ range. The matched cross-size comparison in Table~\ref{tab:base} further shows that both base policies transfer meaningfully, even though $b_1$ performs somewhat better than $b_0$.

\subsection{Reliability--efficiency trade-off}
\label{sec:results_tradeoff}

\begin{figure}[t]
  \centering
  \includegraphics[width=0.85\linewidth]{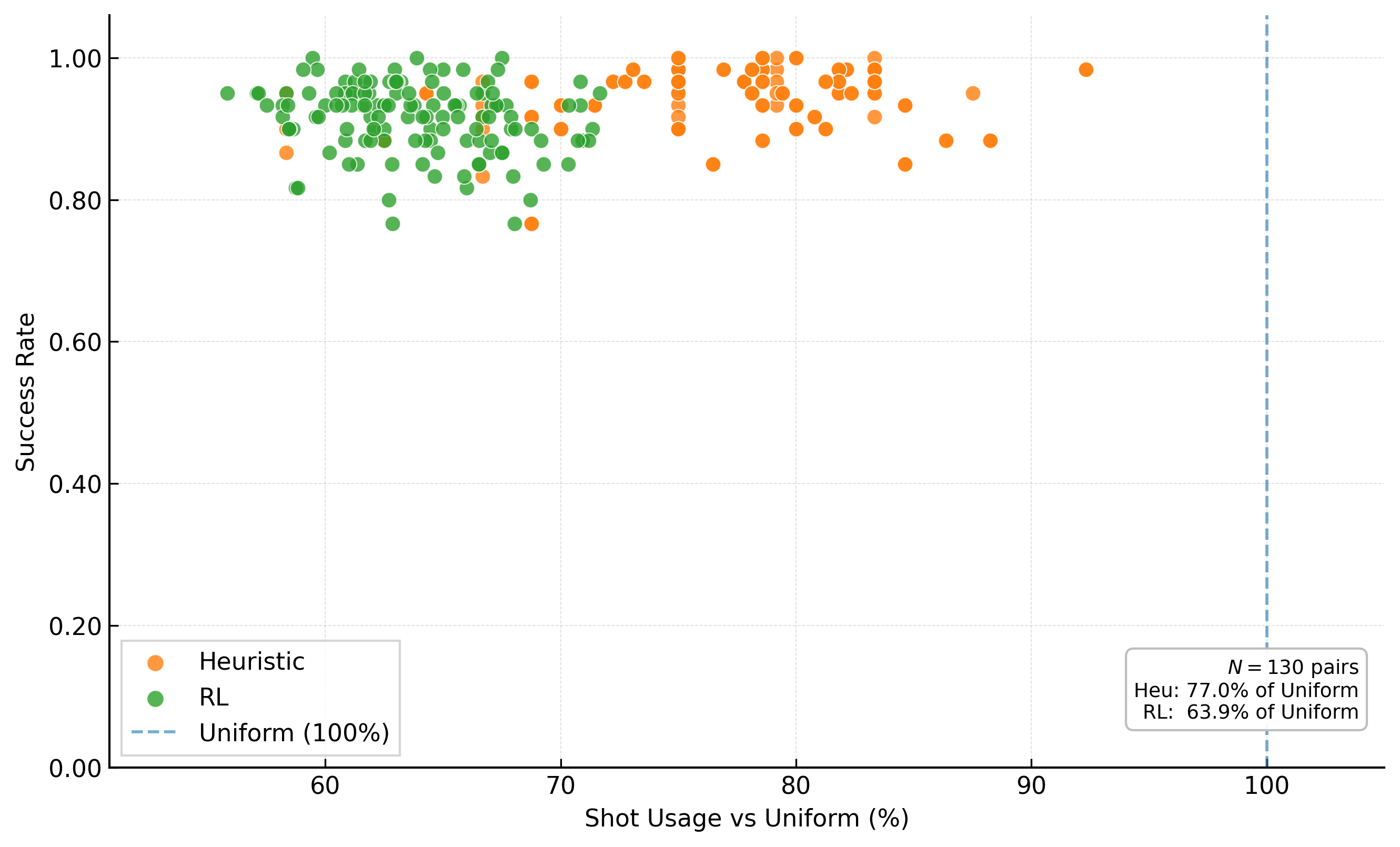}
  \caption{Normalized shot usage ($\widetilde{S}_{\text{method}} / \widetilde{S}_{\text{Uniform}}$) vs.\ success rate for all pairs in the held-out operational subset. Uniform (blue), Heuristic (orange), and RL (green). RL consistently shifts toward lower shot usage while maintaining competitive success rates; the Heuristic occupies an intermediate position.}
  \label{fig:pareto}
\end{figure}

Figure~\ref{fig:pareto} plots normalized shot usage against success rate for all pairs in the held-out operational subset. Relative to Uniform, the RL points are shifted toward lower shot usage while remaining in a high-success regime in the large majority of cases. The Heuristic occupies an intermediate position, with smaller shot savings and slightly higher success rates on average.

The RL policy is trained with a constraint target $p^* = 0.95$, yet the observed success rate on the held-out operational subset is only about 0.91. This gap is not contradictory: the constraint is enforced during training on a designated base instance through the adaptive multiplier $\lambda$, whereas test-time evaluation aggregates many held-out instances of varying difficulty. The held-out results suggest a favorable but nontrivial trade-off: RL gives up a modest amount of reliability relative to Uniform while achieving much larger shot savings, and the resulting ESP ratio remains the best among the tested methods.

To quantify the role of the adaptive constraint in maintaining reliability, we compare the default controller (residual action with adaptive $\lambda$) with an unconstrained variant ($\lambda=0$, all other settings identical) on 22 matched cross-size instances. Table~\ref{tab:sr_floor} reports coverage, defined as the number of matched instance pairs achieving $\mathrm{SR} \geq \tau$, at several reliability floors.

\begin{table}[t]
\centering
\caption{SR-floor coverage comparison on 22 matched cross-size instances. Default denotes the residual controller with adaptive $\lambda$; No Constraint sets $\lambda=0$ throughout training. Coverage is the number of pairs (out of 22) achieving $\mathrm{SR}\geq\tau$.}
\label{tab:sr_floor}
\small
\begin{tabular}{@{}crrr@{}}
\toprule
$\tau$ (SR floor) & Default & No Constraint & $\Delta$ Coverage \\
\midrule
0.90 & 18/22 & 16/22 & $+2$ \\
0.92 & 14/22 &  8/22 & $+6$ \\
0.94 & 13/22 &  5/22 & $+8$ \\
0.95 & 13/22 &  5/22 & $+8$ \\
\bottomrule
\end{tabular}
\end{table}

At $\tau=0.95$, the default controller remains feasible on 13 of 22 instances, compared with 5 for the unconstrained variant, indicating that the adaptive Lagrangian contributes meaningfully to high-reliability coverage. The coverage gap already appears at moderately strict floors and persists through tighter thresholds. This pattern suggests that the adaptive constraint primarily widens the set of instances on which the controller can satisfy strict reliability targets.

\begin{figure}[t]
  \centering
  \includegraphics[width=0.85\linewidth]{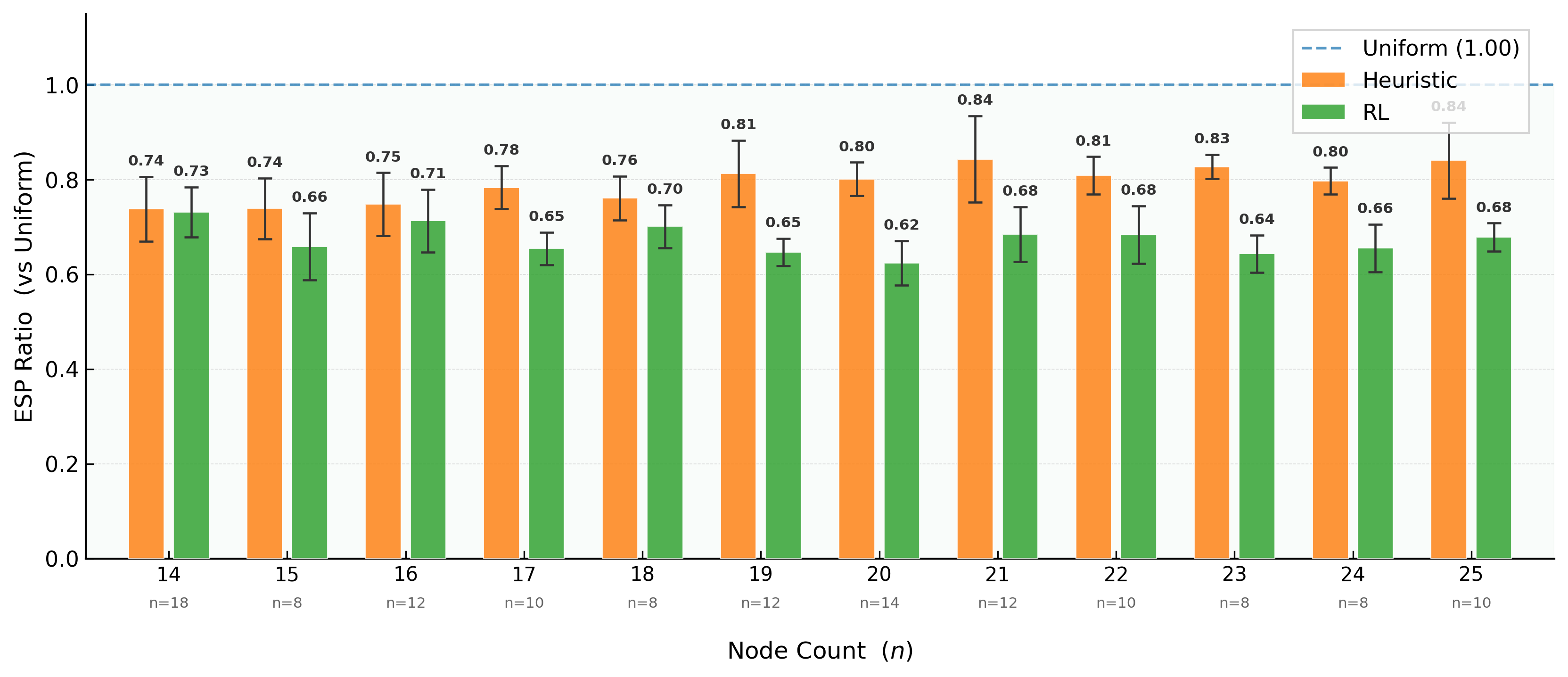}
  \caption{Mean ESP ratio by problem size $n$ for the held-out operational subset. Values below 1.0 indicate lower effective cost per success than Uniform. Error bars show the range across instances at each $n$. Both Heuristic and RL improve over Uniform, with RL achieving a consistently lower ESP ratio.}
  \label{fig:esp_bar}
\end{figure}

\subsection{Sensitivity to filtering and metric choice}
\label{sec:results_filtering}

The main quantitative claims in this paper are based on the held-out operational subset
filtered by $\mathrm{SR}_{\mathrm{Uniform}} \ge 0.90$ after cap calibration. A natural
concern is whether this filtering inflates the reported gains. Table~4 shows that the
qualitative ranking is stable as the inclusion criterion is relaxed: on the held-out
operational subset (130 pairs), RL achieves 36.1\% reduction with an ESP ratio of
0.676, compared with 23.0\% and 0.790 for the Heuristic. Including all held-out pairs
(158 pairs) and then the complete benchmark (164 pairs) changes these numbers only
negligibly. The main conclusions are therefore not an artifact of the filtering boundary.

\begin{table}[t]
\centering
\caption{Filtering sensitivity. The first row applies the primary evaluation protocol ($\mathrm{SR}_{\mathrm{Uniform}} \geq 0.90$, held-out); subsequent rows progressively relax the inclusion criterion. RL reduction and ESP ratio are stable across all subsets.}
\label{tab:filter_sens}
\small
\begin{tabular}{@{}lrcccc@{}}
\toprule
Subset & $N$ & RL Red.\ & RL ESP & Heu.\ Red.\ & Heu.\ ESP \\
\midrule
$\mathrm{SR} \geq 0.90$, held-out            & 130 & 36.1\% & 0.676 & 23.0\%  & 0.790 \\
All held-out pairs                            & 158 & 36.5\% & 0.674 & 22.2\%  & 0.800 \\
Complete benchmark                            & 164 & 36.5\% & 0.674 & 22.3\%  & 0.798 \\
\bottomrule
\end{tabular}
\end{table}

A separate concern is the choice of summary metric. Table~5 reports four shot-reduction
metrics on the held-out operational subset ($N = 130$). The median-based reduction
(36.1\%) and mean-based reduction (36.0\%) are nearly identical for RL, confirming
that the savings are not driven by outlier trials. The tail-sensitive P90 metric still
shows a 27.2\% reduction for RL and 14.2\% for the Heuristic, while the restart-cost
metric $S_{\mathrm{mean}}/\mathrm{SR}$ yields 32.3\% and 20.9\%, respectively. Success
rates on this subset are 0.915 for RL and 0.943 for the Heuristic. Across all four
metrics, RL retains its advantage over the Heuristic.

\begin{table}[t]
\centering
\caption{Metric robustness on the held-out operational subset ($N=130$).
All four metrics confirm RL's advantage. Success rates are 0.915 (RL) and
0.943 (Heuristic) throughout.}
\label{tab:metric_robust}
\small
\begin{tabular}{@{}lcc@{}}
\toprule
Metric & RL Reduction & Heu.\ Reduction \\
\midrule
Median shots (primary)                          & 36.1\% & 23.0\% \\
Mean shots                                      & 36.0\% & 22.9\% \\
P90 (worst-case tail)                           & 27.2\% & 14.2\% \\
Restart cost ($\widetilde{S}_{\mathrm{mean}}/\mathrm{SR}$) & 32.3\% & 20.9\% \\
\bottomrule
\end{tabular}
\end{table}

\subsection{Training dynamics and robustness checks}
\label{sec:results_design}

This subsection provides empirical evidence that the agent learns effectively under the Lagrangian-constrained objective and that the learned policy is robust to the penalty regime.

\begin{figure}[t]
  \centering
  \includegraphics[width=0.95\linewidth]{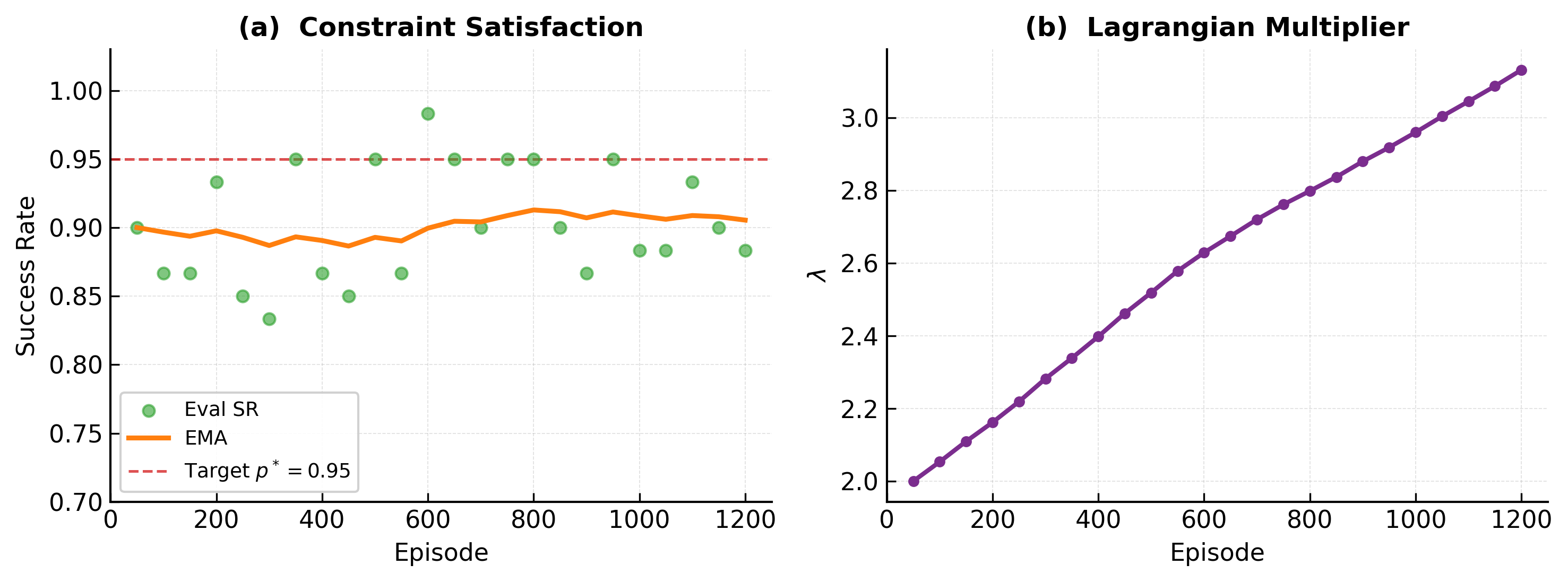}
  \caption{Training diagnostics for the $b_0$ policy ($n{=}14$).
  (a)~Evaluation success rate and its exponential moving average (EMA) remain near the target $p^*{=}0.95$ (dashed line).
  (b)~The Lagrangian multiplier $\lambda$ rises monotonically from 2.0 to ${\sim}3.1$, confirming that the policy sustains feasibility under a progressively tighter constraint.}
  \label{fig:training}
\end{figure}

\paragraph{Training dynamics.}
Figure~\ref{fig:training} shows two training diagnostics for the $b_0$ policy ($n{=}14$). The evaluation success rate remains near the target $p^*{=}0.95$ throughout training (panel~a), while the Lagrangian multiplier $\lambda$ rises steadily from 2.0 to approximately 3.1 (panel~b). Because the reward includes a term $-\lambda(1{-}\sigma)$, the rising $\lambda$ continuously tightens the feasibility constraint; the fact that the success rate is maintained despite this increasingly stringent penalty indicates that the policy learns to reduce shot usage while keeping reliability near the training target. The resulting shot savings are quantified in the held-out evaluation (Table~\ref{tab:aggregate}), where the best-checkpoint policy achieves 36.1\% reduction across 130 unseen evaluation pairs.

\paragraph{Penalty regime sensitivity.}
To assess whether the learned policy is sensitive to the Lagrangian hyperparameters, we trained an aggressive variant with $\lambda_0$ raised fourfold (from 2.0 to 8.0), $\lambda_{\max}$ nearly doubled (from 80 to 150), and the episode budget doubled (from 1{,}200 to 2{,}400), among other changes (full details in Appendix~\ref{app:penalty_sensitivity}). On a matched set of 20 cross-size instances, the aggressive variant achieved a mean shot reduction of 36.9\% with a mean success rate of 0.911, compared with 37.0\% and 0.926 for the standard setting. The near-identical shot reduction despite substantially different penalty regimes indicates that the learned allocation strategy is robust to the constraint-enforcement hyperparameters.

\paragraph{Role of residual structure and adaptive constraint.}
The current controller combines a residual action parameterization with an adaptive Lagrangian constraint. During development, both choices were observed to improve training stability relative to alternatives explored at that stage. A systematic controlled comparison among controller architectures under the final codebase is deferred to future work; the penalty-regime robustness analysis above and the per-step allocation analysis in Section~\ref{sec:discussion} provide partial support for the selected design.
\section{Discussion}
\label{sec:discussion}

\paragraph{Effectiveness of the residual design.}
The residual parameterization provides interpretability and baseline-guided exploration. The Heuristic baseline is not a strawman or a mere initialization device; it is a novel, state-aware allocation rule derived from local indicators of step difficulty. Its own strong performance---achieving approximately 23\% reduction across the benchmark---shows that a compact domain-informed prior already captures a meaningful part of the budget-allocation structure. RL is useful because it learns when this prior should be tightened or relaxed on a step-by-step basis, rather than having to discover the entire allocation policy from scratch.

\begin{figure}[t]
  \centering
  \includegraphics[width=0.95\linewidth]{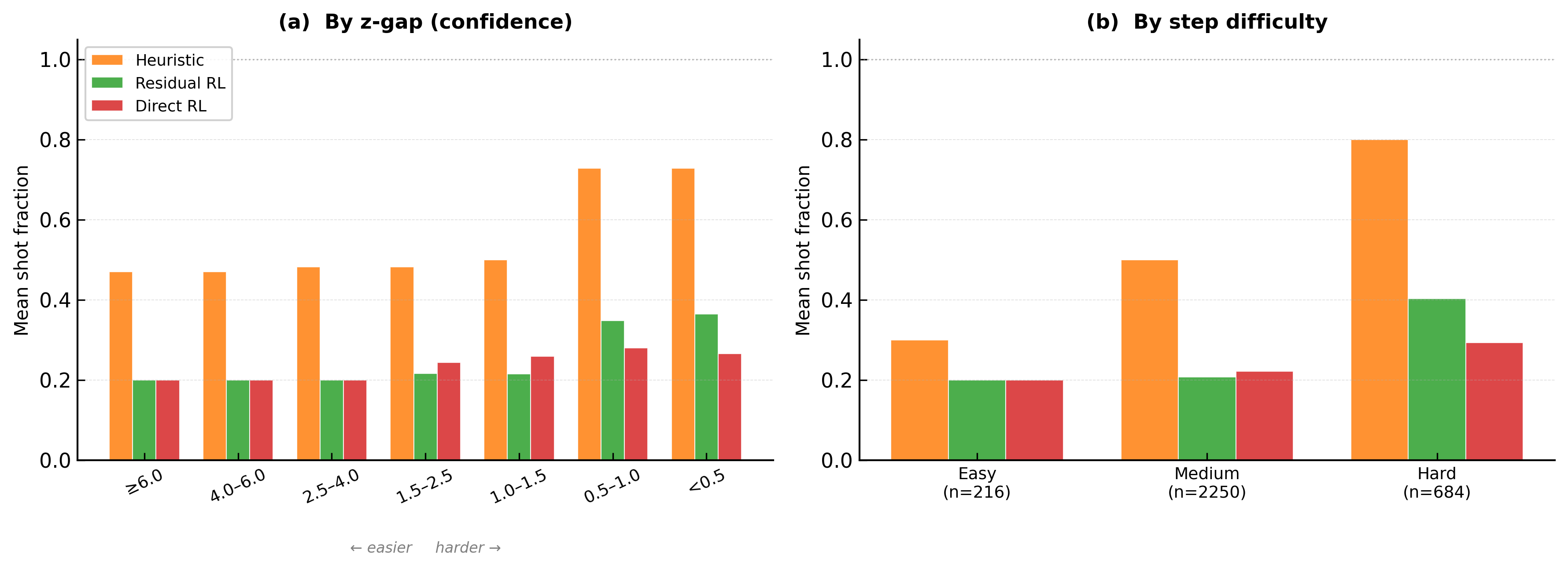}
  \caption{Per-step allocation pattern by step difficulty. Left: mean shot fraction allocated by each controller (Heuristic, Residual RL, Direct RL) as a function of aggregated step difficulty. Right: action distribution across difficulty levels. The residual controller exhibits higher selectivity (hard $-$ easy $= +0.21$) than the direct-action controller ($+0.10$), concentrating additional shots on hard steps while matching the minimum allocation on easy steps.}
  \label{fig:perstep_allocation}
\end{figure}

To characterize what the residual controller learns, we analyze the allocation pattern from the converged Q-tables. Figure~\ref{fig:perstep_allocation} shows the mean shot fraction as a function of z-gap confidence and aggregated step difficulty. The residual controller allocates shots selectively: on easy steps it uses the minimum fraction (0.20), matching the direct-action baseline, but on hard steps it invests approximately twice the budget (0.41 vs.\ 0.20, selectivity $+0.21$). By contrast, a direct-action controller trained under identical conditions converges to near-uniform minimum allocation regardless of difficulty (91\% of states at $\mathrm{frac}=0.20$, selectivity $+0.10$). This indiscriminate reduction is consistent with the lower success rates observed for direct-action policies and helps explain why the residual parameterization enables more effective difficulty-aware allocation. Its relative merit compared with alternative action-space designs warrants further investigation under the final codebase.

\paragraph{Training on hard instances.}
Both base policies are trained on instances that are hard for depth-1 RQAOA (approximation ratio $\leq 0.95$ under Uniform allocation). Hard instances expose a wider range of step difficulties during training---including near-ties in correlations, high conflict ratios, and close candidate edges---providing richer learning signal than easy instances where most steps are unambiguous. This choice is reflected in the base-policy comparison (Table~\ref{tab:base}): $b_1$, trained on the larger and structurally more complex $n{=}20$ instance, achieves higher shot reduction and lower ESP ratio than $b_0$ on the same matched cross-size set, suggesting that a more diverse training environment leads to a more effective allocation policy.

\paragraph{Transfer across problem sizes.}
The cross-size results refer specifically to the held-out Cross-size set of 98 pairs, not to a benchmark that includes Training or Validation categories. The four-dimensional state $(m_t,\zeta_t,\kappa_t,d_t)$ appears to support this transfer because it uses mostly ratio-based or local structural features rather than absolute graph statistics. It is worth noting that the Heuristic itself also transfers to unseen sizes (22.2\% mean reduction on the Cross-size subset), because its rules are defined in terms of the same state features. The contribution of RL is not to enable transfer \emph{per se} but to learn corrections that consistently improve upon the Heuristic's fixed thresholds across the full tested size range. We note, however, that all evaluation instances belong to the same $d$-regular Gaussian-weighted benchmark family, even though not every category is hard-screened; whether these corrections remain effective on structurally different graph classes is an open empirical question.

\paragraph{When does RL help the least?}
The regime in which RL offers the smallest incremental improvement over the Heuristic is same-size evaluation under $b_0$ ($n{=}14$). With only six elimination steps, these instances provide limited room for state-dependent optimization, and the Heuristic already captures most of the available allocation structure (ESP ratio 0.738 vs.\ 0.731 for RL). By contrast, under $b_1$ ($n{=}20$, twelve elimination steps) RL outperforms the Heuristic on all 14 same-size held-out instances (ESP ratio 0.624 vs.\ 0.802). We attribute this pattern to instance complexity rather than to a regime-specific advantage: longer recursion chains expose more steps at which the learned controller can identify and exploit state-dependent opportunities that fixed thresholds miss. From a practical standpoint, this scaling behavior is encouraging because the problems of greatest interest on near-term quantum devices are precisely those at larger sizes, where the RL advantage is most pronounced.

\paragraph{Separation from learned elimination policies.}
Unlike Patel \emph{et al.}\cite{patel2024rlrqaoa}, who learn the elimination policy itself, we keep the elimination rule fixed and optimize only the measurement budget across recursive steps. This isolates the contribution of adaptive shot allocation, preserves the interpretability of the recursive solver, and makes the learned controller easier to deploy as a plug-in replacement for uniform shot allocation. In this sense, our work is complementary to learned elimination-policy methods rather than a competing redesign of the recursive solver. This also distinguishes our setting from optimizer-side shot adaptation in VQE/VQA\cite{kubler2020,arrasmith2020,gu2021,ito2023,sweke2020}, where the controller typically allocates shots during parameter updates rather than along a recursive elimination chain whose measurements change the next reduced instance.

\paragraph{Limitations.}
Several limitations constrain the scope of the present results. First, all experiments are conducted on a noiseless simulator. On real hardware, shot noise interacts with gate errors, readout noise, and decoherence, and the best allocation strategy may shift accordingly\cite{temme2017,kandala2019_error,endo2021}. Second, the study is restricted to depth-1 QAOA. Higher depths may require a richer state representation and a more expressive controller. Third, each base policy is trained on a single designated instance. Although the transfer results are encouraging, multi-instance training may improve robustness. Fourth, the current controller is tabular; substantially larger problems may require function approximation. Fifth, controlled single-variable ablations comparing the residual structure and adaptive constraint against alternatives under the final codebase would strengthen the evidence for these design choices; such ablations are left to future work.
\section{Conclusion}
\label{sec:conclusion}
We presented a two-level framework for dynamic shot allocation in depth-1 RQAOA. A state-aware Heuristic baseline provides an interpretable allocation rule that already achieves approximately 23\% shot reduction, and a residual Double Q-learning controller learns step-by-step corrections that bring the total reduction to 36\% on a held-out benchmark of 130 evaluation pairs over 81 unique instances. The improvement scales with instance complexity and persists on unseen problem sizes, with the Cross-size regime retaining a 36.6\% reduction. Training diagnostics, penalty-regime robustness checks, and per-step allocation analysis support the stability and interpretability of the selected controller configuration.
From a practical standpoint, the learned controller acts as a plug-in replacement for uniform shot allocation in any existing RQAOA pipeline, without modifying the elimination rule or circuit construction. Because the RL advantage grows with problem size, the approach is well positioned for the larger instances that are of primary interest on near-term quantum hardware.
Future directions include extending the framework to higher QAOA depths, where the richer correlation structure may offer additional room for adaptive allocation; validating on noisy hardware, where shot noise interacts with gate and readout errors; multi-instance training, which may improve robustness beyond single-instance policies; and evaluation on diverse graph families beyond the hard $d$-regular weighted instances studied here, including non-regular topologies and broader weight distributions. More broadly, the present results suggest that recursion-level measurement control is a promising and largely unexplored complement to circuit-level and parameter-level optimization in shot-limited quantum algorithms.

\appendix
\section{Implementation Details}
\label{app:impl}
 All quantum circuits are executed with Qiskit Aer using the \texttt{AerSimulator} backend\cite{aleksandrowicz2019}. At each elimination step, the QAOA variational angles $(\gamma, \beta)$ are optimized by a grid search over 48 candidate $\gamma$ values in $[0, 2\pi]$, followed by a SciPy BFGS local refinement on both parameters. Probe shots used to estimate the state features are set to $k_{\mathrm{probe}} = 16$ for $n \leq 16$ and $k_{\mathrm{probe}} = 32$ for $n > 16$.
 Training hyperparameters are fixed across both base policies: learning rate $\alpha = 0.15$, discount factor $\gamma = 0.97$, $\epsilon$-greedy exploration decaying from $1.0$ to $0.02$ at rate $0.995$ per episode, and $1{,}200$ training episodes. The Lagrangian multiplier is initialized at $\lambda_0 = 2.0$, held fixed for a 100-episode warm-up, and subsequently adapted with step size $\mu_\lambda = 1.0$ and ceiling $\lambda_{\max} = 80$. Checkpoint selection retains the policy with the highest validation success rate, breaking ties by lower median total shots and then lower mean total shots.
  \section{Penalty Regime Sensitivity}
\label{app:penalty_sensitivity}
 To assess whether the learned allocation strategy is sensitive to the Lagrangian constraint hyperparameters, we compared the standard training configuration (v3) with an aggressive variant (v4) that simultaneously increases the penalty strength and the training budget. Table~\ref{tab:penalty_config} summarizes the key differences; all other settings (residual action space, state features, action fractions) remain identical.
 \begin{table}[h]
\centering
\caption{Hyperparameter comparison for the penalty regime sensitivity experiment.}
\label{tab:penalty_config}
\small
\begin{tabular}{@{}lrr@{}}
\toprule
Parameter & Standard (v3) & Aggressive (v4) \\
\midrule
$\lambda_0$ & 2.0 & 8.0 \\
$\lambda_{\max}$ & 80 & 150 \\
$\mu_\lambda$ (step size) & 1.0 & 2.0 \\
Warm-up episodes & 100 & 50 \\
Additional fail penalty & 0.0 & 5.0 \\
Training episodes & 1{,}200 & 2{,}400 \\
Evaluation trials & 60 & 100 \\
\bottomrule
\end{tabular}
\end{table}
 Because multiple hyperparameters were changed simultaneously, this comparison is best interpreted as a stress test of the penalty regime rather than a single-variable ablation. Table~\ref{tab:penalty_results} reports test performance on a matched set of 20 cross-size instances evaluated under both configurations.
 \begin{table}[h]
\centering
\caption{Performance comparison on 20 matched cross-size instances under the standard and aggressive penalty regimes.}
\label{tab:penalty_results}
\small
\begin{tabular}{@{}lrr@{}}
\toprule
Metric & Standard (v3) & Aggressive (v4) \\
\midrule
Mean SR & 0.926 & 0.911 \\
Mean shot reduction (\%) & 37.0 & 36.9 \\
\bottomrule
\end{tabular}
\end{table}
 The near-identical shot reduction (37.0\% vs.\ 36.9\%) and comparable success rates despite a fourfold increase in $\lambda_0$, doubled episode budget, and additional failure penalty indicate that the learned policy is robust to substantial changes in the constraint-enforcement regime. During training, $\lambda$ rose to approximately 3.1 under the standard setting and to approximately 12.0 under the aggressive setting, yet the two resulting policies selected similar allocation strategies. This suggests that the final policy is driven primarily by the state features and residual structure rather than by the specific penalty magnitude.
 
\backmatter
 
\bmhead{Acknowledgements}
 
This research was supported by the Quantum Computing based on Quantum Advantage challenge research (RS-2025-08182968) through the National Research Foundation of Korea (NRF) funded by the Korean government (Ministry of Science and ICT (MSIT)).

\section*{Declarations}

\subsection*{Ethics Approval}
Not applicable.

\subsection*{Funding}
This research was supported by the National Research Foundation 
of Korea (NRF), grant number RS-2025-08182968.





\bibliography{sn-bibliography}

@article{lee2025qwmc,
  author  = {Lee, Euimin and Lee, Sangmin and Kim, Shiho},
  title   = {Quantum walk based {Monte Carlo} simulation for photon interaction cross sections},
  journal = {Phys. Rev. D},
  volume  = {111},
  pages   = {116001},
  year    = {2025},
  doi     = {10.1103/PhysRevD.111.116001}
}

@article{ozaeta2022,
  author    = {Ozaeta, Asier and van Dam, Wim and McMahon, Peter L.},
  title     = {Expectation values from the single-layer quantum approximate optimization algorithm on {I}sing problems},
  journal   = {Quantum Science and Technology},
  volume    = {7},
  number    = {4},
  pages     = {045036},
  year      = {2022},
  doi       = {10.1088/2058-9565/ac9f2b}
}

@book{nielsen2010,
  author    = {Nielsen, Michael A. and Chuang, Isaac L.},
  title     = {Quantum Computation and Quantum Information},
  publisher = {Cambridge University Press},
  address   = {Cambridge},
  year      = {2010}
}

@article{preskill2018,
  author  = {Preskill, John},
  title   = {Quantum computing in the {NISQ} era and beyond},
  journal = {Quantum},
  volume  = {2},
  pages   = {79},
  year    = {2018}
}

@article{peruzzo2014,
  author  = {Peruzzo, Alberto and McClean, Jarrod and Shadbolt, Peter and Yung, Man-Hong and Zhou, Xiao-Qi and Love, Peter J. and Aspuru-Guzik, Al{\'a}n and O'Brien, Jeremy L.},
  title   = {A variational eigenvalue solver on a photonic quantum processor},
  journal = {Nat. Commun.},
  volume  = {5},
  pages   = {4213},
  year    = {2014}
}

@article{mcclean2016,
  author  = {McClean, Jarrod R. and Romero, Jonathan and Babbush, Ryan and Aspuru-Guzik, Al{\'a}n},
  title   = {The theory of variational hybrid quantum-classical algorithms},
  journal = {New J. Phys.},
  volume  = {18},
  pages   = {023023},
  year    = {2016}
}

@article{kandala2017,
  author  = {Kandala, Abhinav and Mezzacapo, Antonio and Temme, Kristan and Takita, Maika and Brink, Markus and Chow, Jerry M. and Gambetta, Jay M.},
  title   = {Hardware-efficient variational quantum eigensolver for small molecules and quantum magnets},
  journal = {Nature},
  volume  = {549},
  pages   = {242--246},
  year    = {2017}
}

@article{biamonte2017,
  author  = {Biamonte, Jacob and Wittek, Peter and Pancotti, Nicola and Rebentrost, Patrick and Wiebe, Nathan and Lloyd, Seth},
  title   = {Quantum machine learning},
  journal = {Nature},
  volume  = {549},
  pages   = {195--202},
  year    = {2017}
}

@article{cerezo2021,
  author  = {Cerezo, Marco and Arrasmith, Andrew and Babbush, Ryan and Benjamin, Simon C. and Endo, Suguru and Fujii, Keisuke and McClean, Jarrod R. and Mitarai, Kosuke and Yuan, Xiao and Cincio, Lukasz and Coles, Patrick J.},
  title   = {Variational quantum algorithms},
  journal = {Nat. Rev. Phys.},
  volume  = {3},
  pages   = {625--644},
  year    = {2021}
}

@article{bharti2022,
  author  = {Bharti, Kishor and Cervera-Lierta, Alba and Kyaw, Thi Ha and Haug, Tobias and Alperin-Lea, Sumner and Anand, Abhinav and Degroote, Matthias and Heimonen, Hermanni and Kottmann, Jakob S. and Menke, Tim and Mok, Wai-Keong and Sim, Sukin and Kwek, Leong-Chuan and Aspuru-Guzik, Al{\'a}n},
  title   = {Noisy intermediate-scale quantum algorithms},
  journal = {Rev. Mod. Phys.},
  volume  = {94},
  pages   = {015004},
  year    = {2022}
}

@article{tilly2022,
  author  = {Tilly, Jules and Chen, Hongxiang and Cao, Shuxiang and Picozzi, Dario and Setia, Kanav and Li, Ying and Grant, Edward and Wossnig, Leonard and Rungger, Ivan and Booth, George H. and Tennyson, Jonathan},
  title   = {The variational quantum eigensolver: a review of methods and best practices},
  journal = {Phys. Rep.},
  volume  = {986},
  pages   = {1--128},
  year    = {2022}
}

@article{moll2018,
  author  = {Moll, Nikolaj and Barkoutsos, Panagiotis and Bishop, Lev S. and Chow, Jerry M. and Cross, Andrew and Egger, Daniel J. and Filipp, Stefan and Fuhrer, Andreas and Gambetta, Jay M. and Ganzhorn, Marc and Mezzacapo, Antonio and Salis, Gian and Smolin, John and Tavernelli, Ivano and Temme, Kristan},
  title   = {Quantum optimization using variational algorithms on near-term quantum devices},
  journal = {Quantum Sci. Technol.},
  volume  = {3},
  pages   = {030503},
  year    = {2018}
}

@article{bauer2020,
  author  = {Bauer, Bela and Bravyi, Sergey and Motta, Mario and Chan, Garnet Kin-Lic},
  title   = {Quantum algorithms for quantum chemistry and quantum materials science},
  journal = {Chem. Rev.},
  volume  = {120},
  pages   = {12685--12717},
  year    = {2020}
}

@article{lucas2014,
  author  = {Lucas, Andrew},
  title   = {Ising formulations of many {NP} problems},
  journal = {Front. Phys.},
  volume  = {2},
  pages   = {5},
  year    = {2014}
}

@article{goemans1995,
  author  = {Goemans, Michel X. and Williamson, David P.},
  title   = {Improved approximation algorithms for maximum cut and satisfiability problems using semidefinite programming},
  journal = {J. ACM},
  volume  = {42},
  pages   = {1115--1145},
  year    = {1995}
}

@misc{farhi2014,
  author        = {Farhi, Edward and Goldstone, Jeffrey and Gutmann, Sam},
  title         = {A quantum approximate optimization algorithm},
  year          = {2014},
  eprint        = {1411.4028},
  archivePrefix = {arXiv}
}

@article{hadfield2019,
  author  = {Hadfield, Stuart and Wang, Zhihui and O'Gorman, Bryan and Rieffel, Eleanor G. and Venturelli, Davide and Biswas, Rupak},
  title   = {From the quantum approximate optimization algorithm to a quantum alternating operator ansatz},
  journal = {Algorithms},
  volume  = {12},
  pages   = {34},
  year    = {2019}
}

@article{wang2018,
  author  = {Wang, Zhihui and Hadfield, Stuart and Jiang, Zhang and Rieffel, Eleanor G.},
  title   = {Quantum approximate optimization algorithm for {MaxCut}: a fermionic view},
  journal = {Phys. Rev. A},
  volume  = {97},
  pages   = {022304},
  year    = {2018}
}

@article{zhou2020,
  author  = {Zhou, Leo and Wang, Sheng-Tao and Choi, Soonwon and Pichler, Hannes and Lukin, Mikhail D.},
  title   = {Quantum approximate optimization algorithm: performance, mechanism, and implementation on near-term devices},
  journal = {Phys. Rev. X},
  volume  = {10},
  pages   = {021067},
  year    = {2020}
}

@misc{crooks2018,
  author        = {Crooks, Gavin E.},
  title         = {Performance of the quantum approximate optimization algorithm on the maximum cut problem},
  year          = {2018},
  eprint        = {1811.08419},
  archivePrefix = {arXiv}
}

@misc{farhi_typical2020,
  author        = {Farhi, Edward and Gamarnik, David and Gutmann, Sam},
  title         = {The quantum approximate optimization algorithm needs to see the whole graph: a typical case},
  year          = {2020},
  eprint        = {2004.09002},
  archivePrefix = {arXiv}
}

@misc{farhi_worst2020,
  author        = {Farhi, Edward and Gamarnik, David and Gutmann, Sam},
  title         = {The quantum approximate optimization algorithm needs to see the whole graph: worst case examples},
  year          = {2020},
  eprint        = {2005.08747},
  archivePrefix = {arXiv}
}

@article{marwaha2021,
  author  = {Marwaha, Kunal},
  title   = {Local classical {MAX-CUT} algorithm outperforms $p=2$ {QAOA} on high-girth regular graphs},
  journal = {Quantum},
  volume  = {5},
  pages   = {437},
  year    = {2021}
}

@misc{hastings2019,
  author        = {Hastings, Matthew B.},
  title         = {Classical and quantum bounded depth approximation algorithms},
  year          = {2019},
  eprint        = {1905.07047},
  archivePrefix = {arXiv}
}

@article{bravyi2020symmetry,
  author  = {Bravyi, Sergey and Kliesch, Alexander and Koenig, Robert and Tang, Eugene},
  title   = {Obstacles to state preparation and variational optimization from symmetry protection},
  journal = {Phys. Rev. Lett.},
  volume  = {125},
  pages   = {260505},
  year    = {2020}
}

@article{bravyi2022graphcolor,
  author  = {Bravyi, Sergey and Kliesch, Alexander and Koenig, Robert and Tang, Eugene},
  title   = {Hybrid quantum-classical algorithms for approximate graph coloring},
  journal = {Quantum},
  volume  = {6},
  pages   = {678},
  year    = {2022}
}

@misc{bae2023rqaoa,
  author        = {Bae, Eunice and Lee, Soohhan},
  title         = {Recursive {QAOA} outperforms the original {QAOA} for the {MAX-CUT} problem on complete graphs},
  year          = {2023},
  eprint        = {2211.15832},
  archivePrefix = {arXiv}
}

@article{patel2024rlrqaoa,
  author  = {Patel, Yash J. and Jerbi, Sofiene and B{\"a}ck, Thomas and Dunjko, Vedran},
  title   = {Reinforcement learning assisted recursive {QAOA}},
  journal = {EPJ Quantum Technol.},
  volume  = {11},
  pages   = {1},
  year    = {2024}
}

@article{finzgar2024qiro,
  author  = {Fin{\v{z}}gar, Jonatan Roffe and Kerschbaumer, Andreas and Schuetz, Martin J. A. and Mendl, Christian B. and Katzgraber, Helmut G.},
  title   = {Quantum-informed recursive optimization algorithms},
  journal = {PRX Quantum},
  volume  = {5},
  pages   = {020327},
  year    = {2024}
}

@article{brady2024iterative,
  author  = {Brady, Lucas T. and Hadfield, Stuart},
  title   = {Iterative quantum algorithms for maximum independent set},
  journal = {Phys. Rev. A},
  volume  = {110},
  pages   = {052435},
  year    = {2024}
}

@misc{arrasmith2020,
  author        = {Arrasmith, Andrew and Cincio, Lukasz and Somma, Rolando D. and Coles, Patrick J.},
  title         = {Operator sampling for shot-frugal optimization in variational algorithms},
  year          = {2020},
  eprint        = {2004.06252},
  archivePrefix = {arXiv}
}

@article{kubler2020,
  author  = {K{\"u}bler, Jonas M. and Arrasmith, Andrew and Cincio, Lukasz and Coles, Patrick J.},
  title   = {An adaptive optimizer for measurement-frugal variational algorithms},
  journal = {Quantum},
  volume  = {4},
  pages   = {263},
  year    = {2020}
}

@misc{gu2021,
  author        = {Gu, Andrew and Lowe, Angus and Dub, Pavel A. and Coles, Patrick J. and Arrasmith, Andrew},
  title         = {Adaptive shot allocation for fast convergence in variational quantum algorithms},
  year          = {2021},
  eprint        = {2108.10434},
  archivePrefix = {arXiv}
}

@article{sweke2020,
  author  = {Sweke, Ryan and Wilde, Frederik and Meyer, Johannes and Schuld, Maria and Faehrmann, Paul K. and Meynard-Piganeau, Barthélémy and Eisert, Jens},
  title   = {Stochastic gradient descent for hybrid quantum-classical optimization},
  journal = {Quantum},
  volume  = {4},
  pages   = {314},
  year    = {2020}
}

@misc{ito2023,
  author        = {Ito, Kaito and Mizukami, Wataru and Fujii, Keisuke},
  title         = {Latency-aware adaptive shot allocation for variational quantum algorithms},
  year          = {2023},
  eprint        = {2302.04422},
  archivePrefix = {arXiv}
}

@article{zhao2021,
  author  = {Zhao, Andrew and Rubin, Nicholas C. and Miyake, Akimasa},
  title   = {Measurement reduction in variational quantum algorithms},
  journal = {Phys. Rev. Lett.},
  volume  = {127},
  pages   = {110504},
  year    = {2021}
}

@article{huggins2021,
  author  = {Huggins, William J. and McClean, Jarrod R. and Rubin, Nicholas and Jiang, Zhang and Wiebe, Nathan and Whaley, K. Birgitta and Babbush, Ryan},
  title   = {Efficient and noise resilient measurements for quantum chemistry on near-term quantum computers},
  journal = {npj Quantum Inf.},
  volume  = {7},
  pages   = {23},
  year    = {2021}
}

@article{gokhale2020,
  author  = {Gokhale, Pranav and Angiuli, Olivia and Ding, Yongshan and Gui, Kaiwen and Tomesh, Teague and Suchara, Martin and Martonosi, Margaret and Chong, Frederic T.},
  title   = {Minimizing state preparations in variational quantum eigensolver by partitioning into commuting families},
  journal = {npj Quantum Inf.},
  volume  = {6},
  pages   = {34},
  year    = {2020}
}

@article{izmaylov2019,
  author  = {Izmaylov, Artur F. and Yen, Tzu-Ching and Ryabinkin, Ilya G.},
  title   = {Revising the measurement process in the variational quantum eigensolver: is it possible to reduce the number of separately measured operators?},
  journal = {Chem. Sci.},
  volume  = {10},
  pages   = {3746--3755},
  year    = {2019}
}

@article{verteletskyi2020,
  author  = {Verteletskyi, Vladyslav and Yen, Tzu-Ching and Izmaylov, Artur F.},
  title   = {Measurement optimization in the variational quantum eigensolver using a minimum clique cover},
  journal = {J. Chem. Phys.},
  volume  = {152},
  pages   = {124114},
  year    = {2020}
}

@article{crawford2021,
  author  = {Crawford, Ophelia and van Straaten, Barnaby and Wang, Daochen and Parks, Thomas and Campbell, Earl and Brierley, Stephen},
  title   = {Efficient quantum measurement of {Pauli} operators in the presence of finite sampling error},
  journal = {Quantum},
  volume  = {5},
  pages   = {385},
  year    = {2021}
}

@article{stokes2020,
  author  = {Stokes, James and Izaac, Josh and Killoran, Nathan and Carleo, Giuseppe},
  title   = {Quantum natural gradient},
  journal = {Quantum},
  volume  = {4},
  pages   = {269},
  year    = {2020}
}

@article{spall1998,
  author  = {Spall, James C.},
  title   = {Implementation of the simultaneous perturbation algorithm for stochastic optimization},
  journal = {IEEE Trans. Aerosp. Electron. Syst.},
  volume  = {34},
  pages   = {817--823},
  year    = {1998}
}

@book{sutton2018,
  author    = {Sutton, Richard S. and Barto, Andrew G.},
  title     = {Reinforcement Learning: An Introduction},
  edition   = {2},
  publisher = {MIT Press},
  address   = {Cambridge, MA},
  year      = {2018}
}

@article{watkins1992,
  author  = {Watkins, Christopher J. C. H. and Dayan, Peter},
  title   = {Q-learning},
  journal = {Mach. Learn.},
  volume  = {8},
  pages   = {279--292},
  year    = {1992}
}

@inproceedings{hasselt2010,
  author    = {van Hasselt, Hado},
  title     = {Double {Q}-learning},
  booktitle = {Advances in Neural Information Processing Systems 23},
  editor    = {Lafferty, J. D. and Williams, C. K. I. and Shawe-Taylor, J. and Zemel, R. S. and Culotta, A.},
  publisher = {Curran Associates},
  address   = {Red Hook, NY},
  pages     = {2613--2621},
  year      = {2010}
}

@book{altman1999,
  author    = {Altman, Eitan},
  title     = {Constrained {Markov} Decision Processes},
  publisher = {Chapman \& Hall/CRC},
  address   = {Boca Raton, FL},
  year      = {1999}
}

@inproceedings{achiam2017,
  author    = {Achiam, Joshua and Held, David and Tamar, Aviv and Abbeel, Pieter},
  title     = {Constrained policy optimization},
  booktitle = {Proceedings of the 34th International Conference on Machine Learning},
  series    = {Proceedings of Machine Learning Research},
  publisher = {PMLR},
  address   = {Sydney, Australia},
  pages     = {22--31},
  year      = {2017}
}

@inproceedings{tessler2019,
  author    = {Tessler, Chen and Mankowitz, Daniel J. and Mannor, Shie},
  title     = {Reward constrained policy optimization},
  booktitle = {International Conference on Learning Representations},
  year      = {2019}
}

@article{wauters2020,
  author  = {Wauters, Matteo M. and El-Araby, Emanuele and Amin, Mohammad H.},
  title   = {Reinforcement-learning-assisted quantum optimization},
  journal = {Phys. Rev. Res.},
  volume  = {2},
  pages   = {033446},
  year    = {2020}
}

@inproceedings{yao2020,
  author    = {Yao, Jiahao and Bukov, Marin and Lin, Lin},
  title     = {Policy gradient based quantum approximate optimization algorithm},
  booktitle = {Proceedings of Machine Learning Research},
  volume    = {107},
  pages     = {605--634},
  year      = {2020}
}

@misc{khairy2020,
  author        = {Khairy, Sami and Shaydulin, Ruslan and Cincio, Lukasz and Alexeev, Yuri and Balaprakash, Prasanna},
  title         = {Reinforcement-learning-based variational quantum circuits optimization for combinatorial problems},
  year          = {2020},
  eprint        = {1911.04574},
  archivePrefix = {arXiv}
}

@inproceedings{ostaszewski2021,
  author    = {Ostaszewski, Mateusz and Trenkwalder, Lea M. and Masarczyk, Wojciech and Scerri, Eleanor and Dunjko, Vedran},
  title     = {Reinforcement learning for optimization of variational quantum circuit architectures},
  booktitle = {Advances in Neural Information Processing Systems 34},
  publisher = {Curran Associates},
  address   = {Red Hook, NY},
  pages     = {18182--18194},
  year      = {2021}
}

@article{skolik2023,
  author  = {Skolik, Andrea and Mangini, Stefano and B{\"a}ck, Thomas and Macchiavello, Chiara and Dunjko, Vedran},
  title   = {Robustness of quantum reinforcement learning under hardware errors},
  journal = {EPJ Quantum Technol.},
  volume  = {10},
  pages   = {12},
  year    = {2023}
}

@article{fosel2018,
  author  = {F{\"o}sel, Thomas and Tighineanu, Petru and Weiss, Talitha and Marquardt, Florian},
  title   = {Reinforcement learning with neural networks for quantum feedback},
  journal = {Phys. Rev. X},
  volume  = {8},
  pages   = {031084},
  year    = {2018}
}

@article{niu2019,
  author  = {Niu, Murphy Yuezhen and Boixo, Sergio and Smelyanskiy, Vadim N. and Neven, Hartmut},
  title   = {Universal quantum control through deep reinforcement learning},
  journal = {npj Quantum Inf.},
  volume  = {5},
  pages   = {33},
  year    = {2019}
}

@article{bukov2018,
  author  = {Bukov, Marin and Day, Alexandre G. R. and Sels, Dries and Weinberg, Phillip and Polkovnikov, Anatoli and Mehta, Pankaj},
  title   = {Reinforcement learning in different phases of quantum control},
  journal = {Phys. Rev. X},
  volume  = {8},
  pages   = {031086},
  year    = {2018}
}

@misc{altmann2024,
  author        = {Altmann, Philipp and Stein, Jonas and K{\"o}lle, Michael and B{\u{a}}rligea, Adrian and Gabor, Thomas and Phan, Thomas and Feld, Sebastian and Linnhoff-Popien, Claudia},
  title         = {Challenges for reinforcement learning in quantum circuit design},
  year          = {2024},
  eprint        = {2312.11337},
  archivePrefix = {arXiv}
}

@misc{kuo2021,
  author        = {Kuo, En-Jui and Fang, Yao-Lung L. and Chen, Samuel Yen-Chi},
  title         = {Quantum architecture search via deep reinforcement learning},
  year          = {2021},
  eprint        = {2104.07715},
  archivePrefix = {arXiv}
}

@misc{fosel2021,
  author        = {F{\"o}sel, Thomas and Niu, Murphy Yuezhen and Marquardt, Florian and Li, Liang},
  title         = {Quantum circuit optimization with deep reinforcement learning},
  year          = {2021},
  eprint        = {2103.07585},
  archivePrefix = {arXiv}
}

@misc{aleksandrowicz2019,
  author       = {Aleksandrowicz, Gadi and others},
  title        = {Qiskit: an open-source framework for quantum computing},
  howpublished = {Zenodo},
  doi          = {10.5281/zenodo.2562111},
  year         = {2019}
}

@article{temme2017,
  author  = {Temme, Kristan and Bravyi, Sergey and Gambetta, Jay M.},
  title   = {Error mitigation for short-depth quantum circuits},
  journal = {Phys. Rev. Lett.},
  volume  = {119},
  pages   = {180509},
  year    = {2017}
}

@article{endo2021,
  author  = {Endo, Suguru and Cai, Zhenyu and Benjamin, Simon C. and Yuan, Xiao},
  title   = {Hybrid quantum-classical algorithms and quantum error mitigation},
  journal = {J. Phys. Soc. Jpn.},
  volume  = {90},
  pages   = {032001},
  year    = {2021}
}

@article{kandala2019_error,
  author  = {Kandala, Abhinav and Temme, Kristan and C{\'o}rcoles, Antonio D. and Mezzacapo, Antonio and Chow, Jerry M. and Gambetta, Jay M.},
  title   = {Error mitigation extends the computational reach of a noisy quantum processor},
  journal = {Nature},
  volume  = {567},
  pages   = {491--495},
  year    = {2019}
}

@article{mcclean2018,
  author  = {McClean, Jarrod R. and Boixo, Sergio and Smelyanskiy, Vadim N. and Babbush, Ryan and Neven, Hartmut},
  title   = {Barren plateaus in quantum neural network training landscapes},
  journal = {Nat. Commun.},
  volume  = {9},
  pages   = {4812},
  year    = {2018}
}

@article{cerezo_barren2021,
  author  = {Cerezo, Marco and Sone, Akira and Volkoff, Tyler and Cincio, Lukasz and Coles, Patrick J.},
  title   = {Cost function dependent barren plateaus in shallow parametrized quantum circuits},
  journal = {Nat. Commun.},
  volume  = {12},
  pages   = {1791},
  year    = {2021}
}

@article{bittel2021,
  author  = {Bittel, Lennart and Kliesch, Martin},
  title   = {Training variational quantum algorithms is {NP}-hard},
  journal = {Phys. Rev. Lett.},
  volume  = {127},
  pages   = {120502},
  year    = {2021}
}

@article{willsch2020,
  author  = {Willsch, Dennis and Willsch, Madita and Jin, Fengping and De Raedt, Hans and Michielsen, Kristel},
  title   = {Benchmarking the quantum approximate optimization algorithm},
  journal = {Quantum Inf. Process.},
  volume  = {19},
  pages   = {197},
  year    = {2020}
}

@article{liang2024,
  author  = {Liang, Senwei and Zhu, Linghua and Liu, Xiaolin and Yang, Chao and Li, Xiaosong},
  title   = {Artificial-intelligence-driven shot reduction in quantum measurement},
  journal = {Chem. Phys. Rev.},
  volume  = {5},
  pages   = {041403},
  year    = {2024},
  doi     = {10.1063/5.0219663}
}

\end{document}